\documentclass{PoS}
\usepackage{bm,subfigure,epsfig}

\newcommand{\beq}{\begin{equation}}
\newcommand{\eeq}{\end{equation}}
\newcommand{\bea}{\begin{eqnarray}}
\newcommand{\eea}{\end{eqnarray}}

\title{Graphene: from materials science to particle physics.}
\ShortTitle{Graphene: from materials science to particle physics.}

\author{\speaker{Joaqu\'in E. Drut}
       	\thanks{Present address: Theoretical Division, Los Alamos National Laboratory, Los Alamos, NM, 87545, USA.}\\
       	Department of Physics, The Ohio State University, Columbus, OH 43210--1117, USA\\
       	E-mail: \email{jdrut@mps.ohio-state.edu}}

\author{Timo A. L\"ahde \\
        	Helsinki Institute of Physics and Department of Applied Physics, P. O. Box 14100, \\
	Aalto University, FI-00076 Aalto, Espoo, Finland \\
        	E-mail: \email{timo.lahde@tkk.fi}}

\author{Eero T\"ol\"o \\
        	Department of Applied Physics, P. O. Box 14100, Aalto University, \\ 
	FI-00076 Aalto, Espoo, Finland \\
        	E-mail: \email{eero.tolo@tkk.fi}}

\FullConference{The XXVIII International Symposium on Lattice Field Theory \\
		 June 14-19,2010 \\
		 Villasimius, Sardinia, Italy}

\abstract{Since its discovery in 2004, graphene, a two-dimensional hexagonal carbon allotrope, has generated great interest and 
spurred research activity from materials science to particle physics and {\it vice versa}. In particular, graphene has been found to exhibit 
outstanding electronic and mechanical properties, as well as an unusual low-energy spectrum of Dirac quasiparticles giving rise to 
a fractional quantum Hall effect when freely suspended and immersed in a magnetic field. One of the most intriguing puzzles of 
graphene involves the low-temperature conductivity at zero density, a central issue in the design of graphene-based nanoelectronic
components. While suspended graphene experiments have shown a 
trend reminiscent of semiconductors, with rising resistivity at low temperatures, most theories predict a constant or even decreasing 
resistivity. However, lattice field theory calculations have revealed that suspended graphene is at or near the critical coupling for
excitonic gap formation due to strong Coulomb interactions, which suggests a simple and straightforward explanation for the experimental 
data. In this contribution we review the current status of 
the field with emphasis on the issue of gap formation, and outline recent progress and future points of contact between condensed matter 
physics and Lattice QCD.}

\begin{document}


\section{Introduction}

The recent experimental isolation of single atomic layers of graphite, known as graphene, has provided physicists with a 
novel opportunity to study a system with remarkable electronic and many-body properties, which is
easy to manipulate experimentally~\cite{GeimNovoselov,CastroNetoetal}. Even more recently, the advent of 
experiments utilizing samples of suspended graphene, free from the interference of an underlying substrate~\cite{SuspExp}, 
has provided unprecedented insight into the intrinsic properties of graphene. Among other remarkable discoveries, suspended 
graphene has been found to exhibit a carrier mobility which exceeds that of silicon by an order of magnitude, a fractional quantum
Hall effect which is indicative of strong electron-electron interactions, as well as a markedly non-metallic behavior of the DC conductivity 
at low temperatures.

A central property of graphene is that the low-energy electronic spectrum can be described in terms of two flavors of 
massless, four-component fermionic quasiparticles with linear dispersion~\cite{graphene_quasi}. Indeed, due to the hexagonal 
honeycomb arrangement of the carbon atoms in the graphene lattice, the band structure of graphene exhibits two inequivalent 
(but degenerate) ``Dirac cones'' where the conduction and valence bands touch. Since the energy-momentum relation around a Dirac 
point is linear as in relativistic theories, the low-energy description of graphene bears a certain resemblance to massless 
Quantum Electrodynamics (QED). Nevertheless, an important difference is that the Fermi velocity of the quasiparticles in graphene is as 
low as $v_F^{}\simeq c/300$, whereby the electromagnetic interaction is rendered essentially instantaneous. However, it should be
pointed out that such a value of $v_F^{}$ is actually unusually large from a condensed matter point of view.


Such an unusual band structure (often referred to as semimetallic) accounts fairly well for the observed properties
of graphene sheets deposited on a dielectric substrate. While suspended graphene has recently 
come under intense experimental investigation~\cite{SuspExp}, its spectrum is yet to be computed in a 
controlled fashion. From a theoretical perspective, the challenging feature of suspended graphene lies in the smallness
of the dieletric constant $\epsilon = \epsilon_0^{}$ which, in conjunction with the small value of $v_F^{}$, 
results in a graphene analogue of the fine-structure constant $\alpha_g^{} \gtrsim 1$. At such strong coupling, a dynamical 
transition into a phase fundamentally different from the weakly-coupled semimetallic phase of graphene is a strong 
possibility. In graphene sheets deposited on a substrate, such a transition is effectively inhibited due to the suppression
of $\alpha_g^{}$ by the dielectric.

\section{Low-energy effective theory\label{sec:EffectiveTheory}}

The electronic band structure of graphene close to the Fermi level forms the basis of the low-energy effective theory of 
graphene. This band structure is a reflection of the hexagonal arrangement of the carbon atoms, which can be decomposed
into two triangular sublattices~$A$ and~$B$. This leads to the tight-binding model
\bea
H &=& -t \!\!\!\sum_{\langle i,j \rangle,\sigma=\uparrow,\downarrow}\!\!\!
\left(a_{\sigma,i}^{\dagger} b_{\sigma,j}^{} + h.c.\right)
-t' \!\!\!\!\!\!\sum_{\langle \langle i,j \rangle \rangle,\sigma=\uparrow,\downarrow}\!\!\!\!\!\!
\left(a_{\sigma,i}^{\dagger} a_{\sigma,j}^{} +  b_{\sigma,i}^{\dagger} b_{\sigma,j}^{} + h.c.\right),
\eea
where the operators $a_{\sigma,i}^\dagger (a_{\sigma,i}^{})$ and 
$b_{\sigma,i}^\dagger (b_{\sigma,i}^{})$ create (annihilate) an electron of spin $\sigma$ at location $i$ on sublattices~$A$ and~$B$, 
respectively. The first term (involving $t$) takes into account nearest-neighbor interactions (as first done 
by Wallace in Ref.~\cite{Wallace}), and the second term (involving $t'$) the next-to-nearest neighbor ones. The hopping 
parameters that give an optimal fit to the experimentally determined band structure of graphene 
are $t \simeq 2.8$~eV and $t' \simeq 0.1$~eV~\cite{Reich}. 
The tight-binding model with nearest-neighbor hopping can be generalized to a hexagonal Hubbard model by addition of an
on-site Coulomb repulsion term, which has been recently studied~\cite{Meng} within a Quantum Monte Carlo approach.

We shall follow a somewhat different route based on an Effective Field Theory~(EFT) treatment of 
graphene~\cite{Son,Herbut}, which has the advantage of describing the physics of graphene directly in terms of 
the relevant low-energy degrees of freedom, namely charged massless fermionic quasiparticles. 
The EFT description of graphene has an additional advantage as it allows for the direct study of effects due to the 
unscreened, long-range Coulomb interactions between the quasiparticles. In what follows, we shall formulate a {\it 
continuum} Lagrangian field theory valid at low momenta, much smaller than the inverse 
of the interatomic distance $\sim 1.42$~\AA.


\subsection{Continuum formulation}
\label{subsec:Continuum}

The low-energy EFT of graphene may be derived from a tight-binding or Hubbard model description augmented by a long-range
Coulomb interaction~\cite{Herbut}, yielding a theory of $N_f^{}$ Dirac flavors interacting via an instantaneous Coulomb 
interaction. The action (in Euclidean spacetime) of this theory is
\begin{equation}
S_E^{} = -\sum_{a=1}^{N_f^{}} \int d^2x\,dt \: \bar\psi_a^{} \:D[A_0^{}]\: \psi_a^{} 
+\,\frac{1}{2g^2} \int d^3x\,dt \: (\partial_i^{} A_0^{})^2,
\label{SE}
\end{equation}
where $N_f^{} = 2$ for graphene monolayers, $g^2 = e^2/\epsilon_0^{}$ for graphene in vacuum (suspended graphene), 
$\psi_a^{}$ is a four-component Dirac field in (2+1)~dimensions, $A_0^{}$ is a Coulomb field in (3+1)~dimensions, and
\begin{equation}
D[A_0^{}] = \gamma_0^{} (\partial_0^{} + iA_0^{}) + v_F^{}\gamma_i^{} \partial_i^{}, \quad i=1,2
\end{equation}
where the Dirac matrices $\gamma_\mu^{}$ satisfy the Euclidean Clifford algebra $\{ \gamma_\mu^{}, \gamma_\nu^{}\} = 
2\delta_{\mu\nu}^{}$. 
The four-component spinor structure accounts for quasiparticle excitations of sublattices~$A$ and~$B$ around the two Dirac
points in the band structure~\cite{graphene_quasi,Herbut}. The two Dirac points are identified with the two inequivalent 
representations (with opposite parity) of the Dirac matrices in (2+1)~dimensions. In graphene monolayers, $N_f^{} = 2$ owing 
to electronic spin, while $N_f^{} = 4$ is related to the case of two decoupled graphene layers, interacting solely via the 
Coulomb interaction. Consideration of arbitrary $N_f^{}$ is also useful, given that an analytic 
treatment~\cite{Gonzalez} is possible in the limit $N_f^{} \to \infty$.

The strength of the Coulomb interaction is controlled by $\alpha_g^{} = e^2/(4 \pi v_F^{} 
\epsilon_0^{})$, which is the graphene analogue of the fine-structure constant $\alpha \simeq 1/137$ of QED. It is 
straightforward to show that $\alpha_g^{}$ is the only parameter, by rescaling according to
\beq
t' = v_F^{}t, \quad A_0' = A_0^{}/v_F^{}. 
\label{resc}
\eeq
The action~(\ref{SE}) is invariant under spatially uniform gauge transformations (see Sec. \ref{subsec:gauge}). Notice that 
since the $A_0^{}$ field is (3+1)-dimensional, a four-fermion Coulomb interaction of the form
\beq
\frac{
\bar\psi_a^{}(x) \gamma_0^{} \psi_a^{}(x) \:
\bar\psi_b^{}(x') \gamma_0^{} \psi_b^{}(x')}
{|{\bf x} - {\bf x'}|}
\eeq
is recovered by integrating out $A_0^{}$. Nevertheless, for our purposes the original form of the action (quadratic in the fermions) as 
given in Eq.~(\ref{SE}) is preferable.

A central property of the low-energy EFT is that Eq.~(\ref{SE}) respects a global U$(2 N_f^{})$ chiral 
symmetry under the transformations  
\beq
\psi_a^{} \rightarrow \exp(i \Gamma_j^{} \alpha_j^{})\,\psi_a^{}
\eeq
where the matrices $\Gamma_j^{}$ are the $(2 N_f^{})^2$ hermitian generators of U$(2 N_f^{})$, such that for the 
case of graphene monolayers, the group is U$(4)$. It should be noted that the choice of any particular representation 
for the $\Gamma_j^{}$ is completely arbitrary and is not necessary for any calculational purpose, as all relevant 
information is provided by the Clifford algebra. However, the identification of the spinor degrees of freedom with any 
particular Dirac point and graphene sublattice is dependent on the chosen representation. 
This U$(4)$ chiral symmetry, which strictly speaking is a flavor symmetry, can be spontaneously broken down to 
\mbox{U$(2)\times$U$(2)$}, in which case the excitonic condensate $\langle \bar \psi \psi\rangle$ acquires a non-vanishing 
value, signaling the formation of quasiparticle-hole bound states. The same group 
structure is obtained by adding to Eq.~(\ref{SE}) a parity invariant (Dirac) mass term
\beq
\label{eq:massterm}
\int d^2x\,dt \: m_0^{} \bar\psi_a^{} \psi_a^{},
\eeq
which breaks the symmetry explicitly. For the extended theory with $N_f^{}$ flavors, the symmetry breaking pattern is 
\mbox{U$(2 N_f^{}) \to\,$U$(N_f^{})\times$U$(N_f)$}.
Other symmetry breaking patterns, involving the possibilites of magnetic as well as Cooper-like pairing 
instabilities, have been investigated in Refs.~\cite{Herbut, Khveshchenko2}.


\subsection{Effective action and probability measure}
\label{subsec:Nosignproblem}

The partition function corresponding to Eq.~(\ref{SE}) is given by
\beq
{Z}=
\int{\mathcal D}\!A^{}_0{\mathcal D}\psi^{}{\mathcal D}\bar{\psi}^{} \,
\exp(-S^{}_E[\bar{\psi}^{}_a,\psi^{}_a,A^{}_0]),
\eeq
where it is possible to integrate out the fermionic degrees of freedom, as $S_E^{}$ is quadratic in the $\psi_a^{}$. We thus 
obtain
\beq
{Z}=
\int{\mathcal D}\!A^{}_0 \,
\exp(-S^g_E[A^{}_0]) \, \det(D[A_0^{}])^{N^{}_f}_{},
\eeq
where 
\beq
S^g_E = \frac{1}{2g^2} \int d^3x\,dt \: (\partial_i^{} A_0^{})^2
\label{Sgauge}
\eeq
is the pure gauge part of the action. It is of central importance for the convergence of the Monte Carlo algorithm that the 
above determinant has a definite sign, independently of any particular configuration of the gauge field $A^{}_0$. One way to 
establish this property is to proceed by writing $D[A^{}_0]$ in the form
\beq
D[A^{}_0] = \left(\begin{array}{cc} 
	M[A^{}_0] & 0 \\ 
	0 & -M[A^{}_0] 
	\end{array}\right)
	=
	\left(\begin{array}{cc} 
	M[A^{}_0] & 0 \\ 
	0 & M^\dagger[A^{}_0] 
	\end{array}\right),
\eeq
where 
\beq
M[A^{}_0] = \sigma_0^{} (\partial_0^{} + iA_0^{}) + v_F^{}\sigma_i^{} \partial_i^{}, \quad i=1,2,
\eeq
which entails a specific choice of Dirac $\gamma$-matrices. Furthermore, we note that
$A_0^{}$ is real, and that the Pauli matrices and the momentum operator are hermitian. The 
latter implies $\partial_\mu^\dagger = -\partial^{}_\mu$, and therefore
\beq
\det(D) = \det(M) \det(M^{\dagger}) = |\det(M)|^2 > 0.
\eeq
While this property is not affected by the introduction of a parity invariant mass term such as Eq.~(\ref{eq:massterm}),
the positivity of $\det(D)$ breaks down in the presence of a chemical potential.

The fact that $\det(D)$ is positive definite allows for the definition of an effective action that is purely 
real, given by
\beq
\label{Seff}
S_{\mathrm{eff}}[A_0^{}] = -N_f^{}\ln\det(D[A_0^{}]) + S^g_E[A_0^{}],
\eeq
so that the partition function becomes
\beq
\label{Z_eff}
{Z} = \int{\mathcal D}\!A^{}_0 \,
\exp(-S^{}_{\mathrm{eff}}[A^{}_0]),
\eeq
where $P[A^{}_0] = \exp(-S^{}_{\mathrm{eff}}[A^{}_0]) > 0$ can be interpreted as a positive definite probability measure 
for a Monte Carlo calculation, as outlined in Section~\ref{sec:lattice}.


\subsection{Operator expectation values
\label{subsec:ExpectationValues}}

The expectation value of a given operator $O[\bar \psi,\psi]$ dependent on the fermion fields can be calculated by taking 
functional derivatives of the generating functional
\beq
{Z[\bar\eta, \eta]} =
\int{\mathcal D}\!A^{}_0 {\mathcal D}\psi^{}{\mathcal D}\bar{\psi} 
\:\exp(-S^{}_{E}[A^{}_0,\bar\psi, \psi, \bar\eta, \eta]),
\eeq
where source terms have been added to the original action according to
\beq
S_E^{} [A^{}_0,\bar\psi, \psi, \bar\eta, \eta] = S_E^{}[A^{}_0,\bar\psi, \psi] 
\,+\int d^2x \, dt \, (\bar\psi \eta + h.c.),
\eeq
such that the modified effective gauge action is a functional of $A_0^{}$ as well as of 
the sources $\eta, \bar\eta$. It is again possible to integrate out the fermionic degrees of freedom and take 
functional derivatives with respect to the sources in the resulting expression
\beq
{Z[\bar\eta, \eta]} \propto \int{\mathcal D}\!A^{}_0 \,
\exp(-S^{}_{\mathrm{eff}}[A^{}_0])
\:\exp\left(-\int d^2x \, dt \,\bar\eta D^{-1}[A_0^{}] \eta\right),
\eeq
which makes it possible to obtain expectation values in terms of a path integral over $A_0^{}$ only. While this 
procedure is completely general, it is possible to employ a slightly different approach in order to facilitate the 
computation of the chiral condensate and susceptibility.

The chiral condensate $\sigma$, which is the order parameter of the semimetal-insulator phase transition in graphene, 
is defined by
\beq
\sigma \equiv \langle \bar{\psi}_b^{} \psi_b^{} \rangle,
\eeq
where the fermion fields are evaluated at the same space-time point. It is useful to note that the mass $m_0^{}$ plays the 
r\^ole of a source, coupled to $\bar\psi_b^{} \psi_b^{}$. The expectation value of this operator can therefore be obtained 
by first differentiating the partition function with respect to $m_0^{}$ and dividing by the volume, giving 
\beq
\sigma = \frac{1}{V Z}\int{\mathcal D}\!A_0^{}{\mathcal D}\psi^{}{\mathcal D}\bar{\psi}
\int d^2x\,dt \, \bar{\psi}_b^{}(x,t) {\psi}_b^{}(x,t) \, \exp(-S_E^{}) 
= \frac{1}{V}\frac{\partial \ln {Z}}{\partial m_0^{}},
\eeq
where $\sigma$ has been defined as an average over the lattice volume occupied by the fermions.
On the other hand, once the 
fermions have been integrated out, the derivative with respect to $m_0^{}$ yields
\beq
\sigma =
\frac{1}{V Z}\int{\mathcal D}\!A_0^{} \, \mathrm{Tr}(D^{-1}[A_0^{}]) \, 
\exp(-S_{\mathrm{eff}}[A_0^{}])
= \frac{1}{V}\left\langle \mathrm{Tr}(D^{-1}[A^{}_0]) \right\rangle,
\eeq
where the identities
\beq
\det(D[\lambda]) = \exp(\mathrm{Tr}(\log(D[\lambda])), \quad
\frac{\partial\det(D[\lambda])}{\partial \lambda} = 
\det(D[\lambda]) \, \mathrm{Tr}\left(D^{-1}[\lambda] \, \frac{\partial D}{\partial \lambda} \right),
\eeq
have been used. The chiral susceptibility $\chi_l^{}$ may be found by taking one more derivative with respect to $m_0^{}$, 
giving
\beq
\chi_l^{}  \equiv \frac{\partial \sigma}{\partial m_0^{}} = \frac{1}{V} \left[ 
\left\langle \mathrm{Tr}^2(D^{-1}) \right\rangle - 
\left\langle \mathrm{Tr}(D^{-2}) \right\rangle - 
\left\langle \mathrm{Tr}(D^{-1}) \right\rangle^2 
\right ],
\eeq
which is expected to diverge at a second-order phase transition, and may also 
yield constraining information on the universal critical exponents of the transition.


\section{Graphene on the lattice}
\label{sec:lattice}

In this section we formulate the lattice version of Eq.~(\ref{SE}) following Refs.~\cite{DL_PRL,DL_PRB}. It should be 
noted in this context that a closely related lattice model of the strong-coupling limit of graphene has been considered
in Ref.~\cite{graphene_Hands}.
We begin by discretizing the pure gauge sector, where 
the requirement of gauge invariance implies the use of ``link variables'' to represent the gauge degrees of freedom. The 
``staggered'' discretization of the fermionic sector is then outlined, as it is the preferred choice to 
represent fermions with chiral symmetry at finite lattice spacing~\cite{Kogutetal,Gockeleretal}.
Throughout this paper, the lattice spacing is set equal to unity, and thus all dimensionful quantities should be regarded as 
expressed in units of the lattice spacing.

\subsection{Gauge invariance and link variables}
\label{subsec:gauge}

The pure gauge part of the Euclidean action, Eq.~(\ref{Sgauge}), can be
thought of as the non-relativistic limit of the Lorentz-invariant form $\frac{1}{4} 
F_{\mu\nu}^{} F^{\mu\nu}_{}$ where $F_{\mu\nu}^{} = \partial_\mu^{} A_\nu^{} - \partial_\nu^{} A_\mu^{}$, such that
\beq
F_{\mu\nu}^{} F^{\mu\nu}_{} = F_{0j}^{} F^{0j}_{} + F_{ij}^{} F^{ij}_{} + F_{i0}^{} F^{i0}_{}
= 2 F_{0j}^{} F^{0j}_{} = 2 (\partial_j^{} A_0^{})^2,
\eeq
where we have used $F_{ij}^{} = 0$ (no magnetic field) and $\partial_0^{} A_j^{} = 0$ (no electric field induction by a 
magnetic field), valid in the non-relativistic limit ($v_F^{} \ll c$). Thus, for graphene the only non-vanishing 
contribution is the electric field $E_j^{} = -\partial_j^{} A_0^{}$, which represents the instantaneous Coulomb interaction 
between the quasiparticles.

The gauge action~(\ref{Sgauge}) is invariant under the time-dependent, spatially uniform gauge transformations
\beq
A_0^{} \rightarrow A_0^{} + \alpha(t), \quad
\psi \rightarrow \exp\left\{i \int_0^t dt' \alpha(t') \right\} \psi,
\eeq
where $\alpha(t)$ is a function of time only. Thus, in spite of its apparent simplicity, the effective theory of graphene 
possesses a truly local gauge invariance, which should be respected by the lattice action. To this end, one introduces 
temporal link variables
\beq
U_{0,{\bf n}}^{} = U_{\bf n}^{} \equiv \exp\left(i\theta_{\bf n}^{}\right),
\eeq
where $\theta_{\bf n}^{}$ is the dimensionless lattice gauge field evaluated at the lattice point ${\bf n} = 
(n_0,n_1,n_2,n_3)$. The spatial link variables
\beq
U_{i,{\bf n}}^{} = 1
\eeq
are set to unity. It is convenient to express the discretized version of Eq.~(\ref{Sgauge}) in terms of 
``plaquette'' variables, defined by
\beq
U_{\mu\nu,{\bf n}}^{} = 
U_{\mu,{\bf n}}^{} U_{\nu,{\bf n} + {\bf e}_\mu^{}}^{} 
U^\dagger_{\mu,{\bf n} + {\bf e}_\nu^{}} U^\dagger_{\nu,{\bf n}},
\eeq
where, in the present case of a pure Coulomb interaction, the only non-trivial components are $U_{0i}^{}$ and $U_{i0}^{}$. 
Those plaquette components then correspond to the discretized formulation of the electric field. The remaining components
corresponding to the magnetic field are equal to unity. These statements can be summarized in the expression
\beq
U_{\mu\nu,{\bf n}}^{} = 
\delta_{\mu 0}^{} \delta_{\nu i}^{} \, U_{\bf n}^{} U^{\dagger}_{{\bf n}+{\bf e}_i^{}}
+ \,\delta_{\nu 0}^{} \delta_{\mu i}^{} \, U^{\dagger}_{\bf n} U_{{\bf n}+{\bf e}_i^{}}^{}
+ \,\delta_{\mu 0}^{} \delta_{\nu 0}^{} + \delta_{\mu i}^{} \delta_{\nu j}^{}.
\eeq

In terms of the gauge link variables and plaquettes, the discretized gauge action corresponding to Eq.~(\ref{Sgauge})
is given by~\cite{Rothe}
\beq
S^g_E = \beta \sum_{\bf n} \sum_{\mu>\nu}{\left [1 -
\frac{1}{2}\left(U_{\mu\nu,{\bf n}}^{} + U^{\dagger}_{\mu\nu,{\bf n}}\right) \right]},
\label{Sgauge2}
\eeq
where $\beta = 1/g^2$, such that $\beta \to v_F^{}/g^2$ when the rescaling of Eq.~(\ref{resc}) is applied. In 
Eq.~(\ref{Sgauge2}), the only non-vanishing contributions arise from the terms with 
$(\mu,\nu) = (1,0);(2,0);(3,0);(2,1);(3,1)$ and $(3,2)$. Eq.~(\ref{Sgauge2}) may be simplified to
\beq
S^g_{E,C} = \beta \sum_{\bf n} {\left [3 - \sum^3_{i=1} \Re 
\left(U_{\bf n}^{} U^\dagger_{{\bf n}+{\bf e}_i^{}}\right) \right]},
\label{SgaugeC}
\eeq
where $\Re(x)$ denotes the real part of $x$.
Eq.~(\ref{SgaugeC}) is referred to as the compact formulation, which has been found to pose problems related to
spurious monopole condensation in QED and related theories~\cite{compact}. On the other hand, the non-compact
formulation, which is obtained from Eq.~(\ref{SgaugeC}) by expanding 
$\Re(U_{\bf n}^{} U^\dagger_{{\bf n}+{\bf e}_i^{}})$ to second order in $\theta$,
\beq
\Re\left(U_{\bf n}^{} U^\dagger_{{\bf n}+{\bf e}_i^{}}\right) = 
1 - \frac{1}{2}\left ( \theta_{{\bf n}+{\bf e}_i^{}}^{} - \theta_{\bf n}^{} 
\right)^2 + \:\ldots
\eeq
giving
\beq
S^g_{E,N} = \frac{\beta}{2} \sum_{\bf n} {\sum^3_{i=1} \left ( \theta_{{\bf n}+{\bf e}_i^{}}^{} - 
\theta_{\bf n}^{} \right)^2 },
\label{SgaugeN}
\eeq
is free from such problems~\cite{Gockeleretal,Noncompact} and allows for a realistic continuum limit. 


\subsection{Staggered fermions}
\label{subsec:staggeredfermions}

While the discretization of the gauge sector is relatively straightforward, the inclusion of dynamical fermions on the 
lattice is a notoriously difficult problem. One of the main issues when simulating fermions on the lattice is the so-called 
doubling problem (for an overview, see Ref.~\cite{Rothe}, Chapter 4). This problem is related to the chiral invariance of 
the fermionic sector, and arises due to the appearance of multiple (unwanted) zeros in the inverse propagator. In other 
words, one is simulating more fermion flavors than expected, the exact number being dependent on the dimensionality of the 
theory. There exists a number of ways to avoid the doubling problem, but all of them break chiral 
invariance in one way or the other, a fact encoded in the Nielsen-Ninomiya theorem~\cite{NielsenNinomiya}. The 
solution we have chosen for our simulations of graphene is the ``staggered'' discretization 
of Ref.~\cite{Kogut-Susskind}. This choice is advantageous for the study of spontaneous chiral symmetry breaking in graphene, as 
it yields the correct number of degrees of freedom while (partially) preserving chiral symmetry. The major drawback of 
staggered fermions is that the full chiral symmetry is restored only in the continuum limit, a fact referred to as ``taste symmetry
breaking''.

In order to discretize the fermionic sector of Eq.~(\ref{SE}) in a way amenable to computer simulations, a number 
of choices need to be made. As a first step, fermions are integrated out, and the 
partition function is written purely in terms of the gauge field, Eq.~(\ref{Z_eff}). The fermions are then represented 
exclusively through $\det(D)$. One can then attempt to compute the determinant exactly for a 
given $\theta$ configuration, which is feasible due to the low dimensionality of the problem,  Alternatively, one may 
rewrite $\det(D)$ in terms of a path integral over complex scalar fields referred
to as pseudofermions, as is common in Lattice QCD.

It has been shown in Ref.~\cite{BurdenBurkitt} that for each staggered flavor one recovers, in the continuum limit, 
two four-component Dirac flavors. Thus, by retaining one staggered flavor, it is possible to have exactly eight continuum 
fermionic degrees of freedom, which is the correct number for graphene. The action of a single staggered flavor
is given by
\beq
S^f_E[\bar{\chi},\chi,\theta] = 
-\sum_{{\bf n},{\bf m}} 
{\bar\chi}_{\bf n}^{} \, K_{{\bf n},{\bf m}}^{}[\theta] \,
{\chi}_{\bf m}^{},
\eeq
where the staggered Dirac operator is
\beq
K_{{\bf n},{\bf m}}^{}[\theta] =
\frac{1}{2}(\delta_{{\bf n}+{\bf e}_0^{},{\bf m}}^{} \, U_{\bf n}^{} - 
\delta_{{\bf n} - {\bf e}_0^{},{\bf m}}^{} \, U^{\dagger}_{\bf m})
+ \frac{v_F^{}}{2} \sum_i \eta^i_{\bf n} (\delta_{{\bf n}+{\bf e}_i^{},{\bf m}}^{} - 
\delta_{{\bf n}-{\bf e}_i^{},{\bf m}}^{})
+ m_0^{} \delta_{{\bf n},{\bf m}}^{},
\label{Kstag}
\eeq
where the phase factors $\eta$ arise from the spin-diagonalization of the Dirac matrices~\cite{KawamotoSmit}.
The operator $K$ thus replaces $D$ in all expressions for the probability, chiral condensate and susceptibility that were 
derived in the previous sections. As expected from the Nielsen-Ninomiya theorem, the staggered lattice action does not 
retain the full U$(4)$ chiral symmetry of the original graphene action at finite lattice spacing. As shown in 
Ref.~\cite{BurdenBurkitt}, only a subgroup U$(1)\times$U$(1)$ remains upon discretization. Spontaneous condensation of 
$\bar\chi \chi$, or equivalently the introduction of a parity invariant mass term, reduces this symmetry to U$(1)$.

Finally, it should be pointed out that the situation 
concerning graphene is unusually favorable, in the sense that the staggered formalism fortuitously provides the 
correct number of fermionic degrees of freedom, as $N_f^{} = 2$ for graphene monolayers. In general, staggered fermions 
provide only a compromise solution in the sense that some degree of chiral symmetry is preserved, at the price of retaining 
some of the doubling originally present in the discretized fermion action. Indeed, if the case of $N_f^{} = 1$ were to be 
simulated, it would be necessary to resort to the controversial ``rooting'' trick~\cite{rooting}, whereby 
the desired number of continuum flavors is restored by taking the appropriate root of the Dirac operator.


\section{Results for $\beta_c^{}$}

\begin{figure}[t]
\epsfig{file = 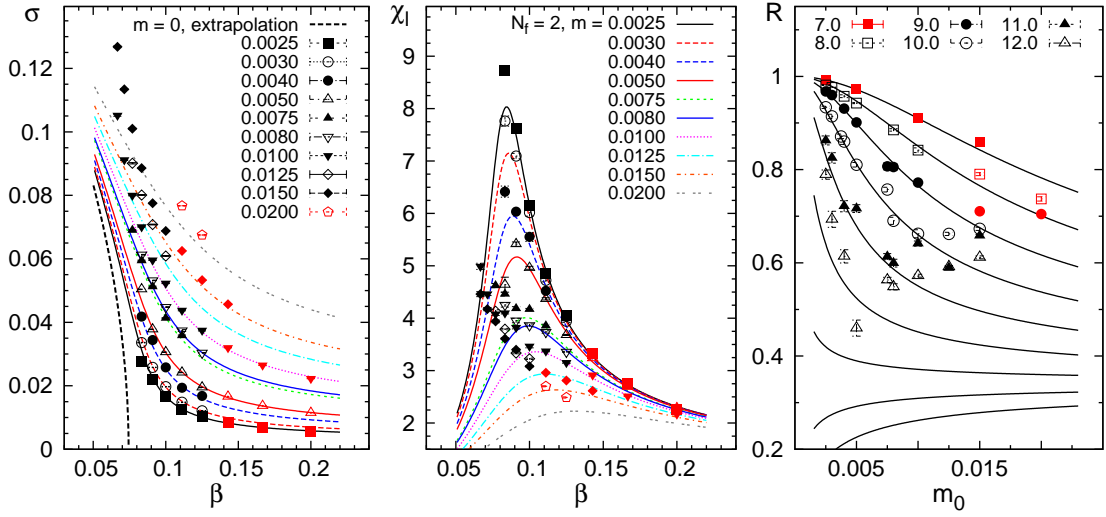, width = \columnwidth}
\caption{Result of a simultaneous fit to $\sigma$ (left panel), $\chi_l^{}$ (middle panel) and $R$ (right panel). 
Red symbols indicate data for $N_x^{} = 28$, the remaining ones are for $N_x^{} = 32$. The datapoints for $R$ are grouped
according to $\beta^{-1}$. The fit range has been 
restricted such that data with $m_0^{} > 0.005$ and $\beta < 0.1$ are excluded, as data at large $m_0^{}$ and small
$\beta$ have substantial finite volume effects. The optimal parameters are $\beta_c^{} = 0.0774(2)$ and 
$\delta = 2.93(2)$, with $b = 1.0$ fixed. The data satisfy $b = 1$ very accurately. The error bars on the datapoints are
obtained via the Jackknife method~\cite{jackknife}.
}
\end{figure}

The results presented in this work for the non-compact gauge action
are partly based on the reanalysis of extant data from Ref.~\cite{DL_PRL}
and partly on new data on larger lattices. These datasets correspond to $N_t^{} = N_x^{} = 28$ and $32$,
with the extent of the bulk dimension $N_b^{}$ set at either $12$ or $32$. We find that finite-size effects are
most pronounced as functions of $N_t^{}$ and $N_x^{}$ at small $\beta$, while the effect of $N_b^{}$ on the
results is negligible. Our analysis proceeds by first determining the condensate $\sigma$, susceptibility $\chi_l^{}$ and
$R \equiv \chi_l^{} \, m_0^{}/\sigma$ as a function of $\beta$ and $m_0^{}$. The second step of the analysis consists of fitting an
equation of state~(EOS) to $\sigma$ and $\chi_l^{}$ in order to obtain estimates of the critical coupling
$\beta_c^{}$ and the critical exponents. Our EOS is of the form
\begin{equation}
m_0^{}X(\beta) = Y(\beta)\,\sigma^b_{} + \sigma^\delta_{},
\label{EOS}
\end{equation}
where the exponent $b \equiv \bar\beta - 1/\delta$. This form has been used previously in the context of
Lattice QED in (3+1)~dimensions~\cite{Gockeler_EOS}.
Here, we have referred to the critical exponent using
the notation $\bar\beta$ to avoid confusion with the inverse coupling $\beta$. The dependence on
the critical coupling $\beta_c^{}$ enters through the expansions
\begin{equation}
X(\beta) = X_0^{} + X_1^{}\left(1-\frac{\beta}{\beta_c^{}}\right) 
+ \ldots,
\quad\quad
Y(\beta) = Y_1^{}\left(1-\frac{\beta}{\beta_c^{}}\right)
+ \ldots,
\end{equation}
where terms up to $X_1^{}$ and $Y_1^{}$ have been retained. Higher-order terms were found
to have a very small effect on the analysis, and have thus been discarded.

We find that a simultaneous fit to all data for $\sigma$, $\chi_l^{}$ and $R$ yields the most stable results. It is
also necessary to carefully consider finite-volume effects and the impact they have on the fit parameters. We have
therefore excluded datapoints with $m_0^{} > 0.005$ and $\beta < 0.1$. In this way, we
find consistency with mean-field exponents, $\delta = 3$ and $b = 1$. Including data at lower $\beta$ and larger $m_0^{}$
suggests $\delta \sim 2.2$, however such fits have a much higher $\chi^2_{}$ and compare unfavorably with the data 
on $R$. We find that $N_x^{} = 28$ and $N_x^{} = 32$ give consistent results using the restricted
dataset.


\section{Experimental situation}

We now turn to the question whether experiments which measure the DC conductivity of suspended graphene provide any evidence
for semiconducting behavior which would follow naturally from the excitonic gap scenario. While a full LMC calculation of the conductivity
is not yet available, a simplified analysis in terms of a Kubo description of gapped quasiparticles has recently been given in Ref.~\cite{DL_gap},
where the data of Ref.~\cite{Bolotin_SG} on the suspended graphene devices ``S1'', ``S2'' and ``S3'' were analyzed in terms of the
expression $\sigma \equiv \sigma_q^{} + \sigma_{bg}^{}$. Here $\sigma_q^{}$ is the quasiparticle contribution intrinsic to graphene, 
while the ``background'' component $\sigma_{bg}^{}$ is device-dependent.

The Hamiltonian describing Dirac quasiparticles with a gap $\Delta$ and Fermi
velocity $v_F^{}\simeq c/300$ is given by
$H \equiv \sigma_1^{}v_F^{}k_1^{} + \sigma_2^{}v_F^{}k_2^{} + \sigma_3^{}\Delta/2$,
where the $\sigma_i$ are Pauli matrices. The contribution $\sigma_q^{}$ of the Dirac 
quasiparticles to the DC conductivity of a graphene monolayer is then
\begin{eqnarray}
\sigma_q^{} &\equiv& \frac{4e^2}{h} \frac{\pi}{2}
\int_{-\infty}^{\infty} d\epsilon\, \int_{\Delta/2}^{\infty} d\xi\,\xi
\mathcal T_\omega^{}(\xi,\epsilon)
\:\: \frac{f(\beta\epsilon-\frac{\beta\omega}{2}-\beta\mu) 
- f(\beta\epsilon+\frac{\beta\omega}{2}-\beta\mu)}{\omega},
\label{qp}
\end{eqnarray}
where $\beta \equiv 1/k_B^{}T$, the Fermi function is given by
$f(x) = 1/(1+\exp(x))$, $\mu$ denotes the chemical potential and
the factor of~$4$ accounts for the spin and valley degrees of freedom. Then
\begin{eqnarray}
\label{Amplitude}
\mathcal T_\omega^{}(\xi,\epsilon) &=& \frac{\xi^2+\Delta^2/4}{\xi^2} 
\left[
\delta_\eta^{}\left(\xi+\epsilon+\frac{\omega}{2}\right)
\delta_\eta^{}\left(\xi-\epsilon+\frac{\omega}{2}\right) +
\delta_\eta^{}\left(\xi+\epsilon-\frac{\omega}{2}\right)
\delta_\eta^{}\left(\xi-\epsilon-\frac{\omega}{2}\right) 
\right] 
\nonumber \\
&+& \frac{\xi^2-\Delta^2/4}{\xi^2}
\left[
\delta_\eta^{}\left(\xi-\epsilon-\frac{\omega}{2}\right)
\delta_\eta^{}\left(\xi-\epsilon+\frac{\omega}{2}\right) +
\delta_\eta^{}\left(\xi+\epsilon+\frac{\omega}{2}\right)
\delta_\eta^{}\left(\xi+\epsilon-\frac{\omega}{2}\right) 
\right], \nonumber \\
\end{eqnarray}
where $\eta$ is the scattering rate of the quasiparticles, which can be
accounted for~\cite{Ziegler} by broadening the delta functions according to $\pi\delta_\eta^{}(x) \equiv 
\eta/(x^2+\eta^2)$. In the DC limit, the integral over $\xi$ in Eq.~(\ref{qp}) yields
\beq
\label{Amplitude_DC}
\int_{\Delta/2}^{\infty}\! d\xi\,\xi\,\mathcal T_0^{}(\xi,\epsilon) = 
\frac{1}{2\pi} - \frac{\Delta^2\!-4|z|^2}{16 \pi \epsilon \eta} \arg \left(\Delta^2 \!- 4 z^2\right), 
\eeq
where $z\equiv\epsilon + i\eta$. 


\newsavebox{\tempbox}
\sbox{\tempbox}{
\epsfig{file=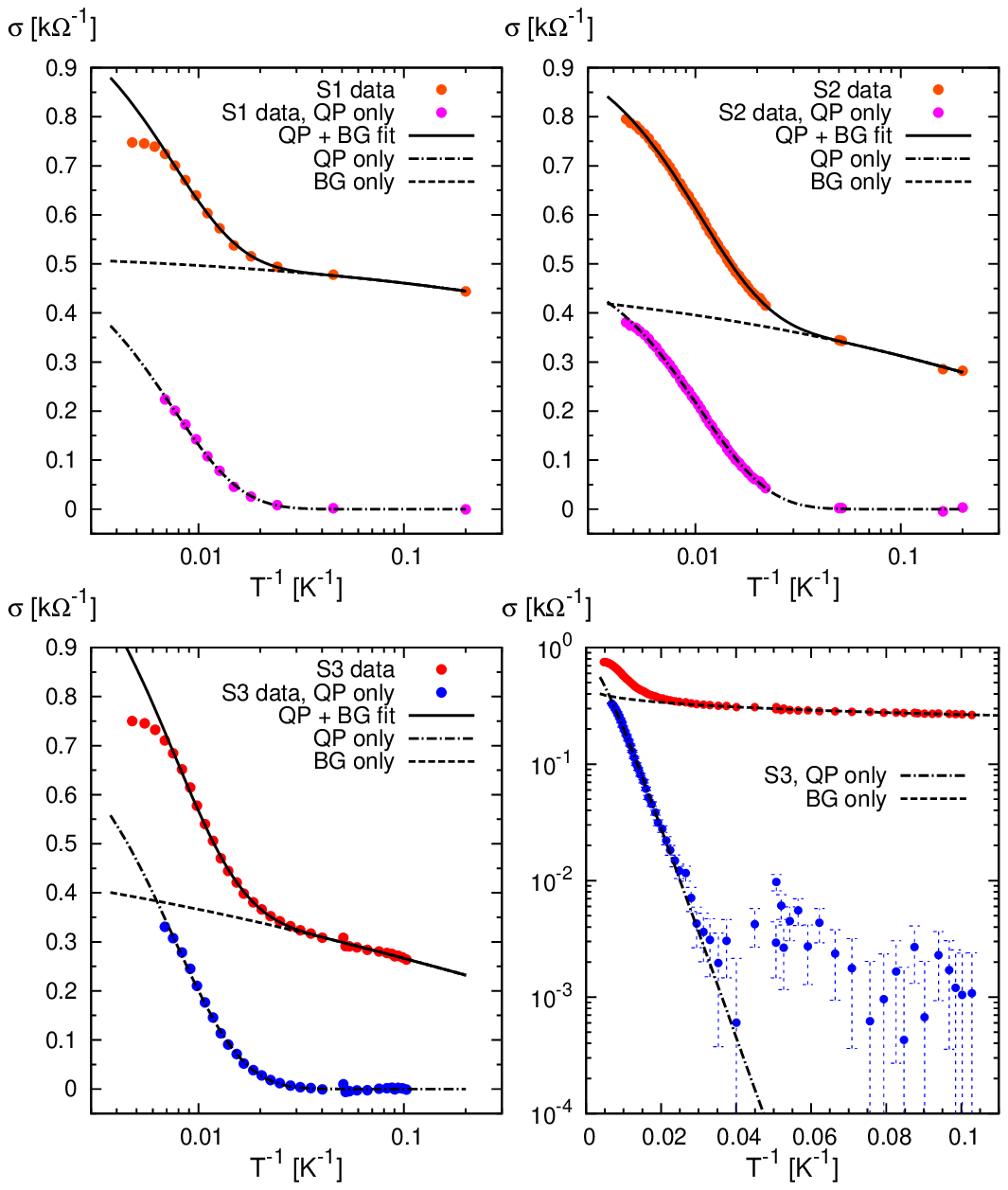, width=.48\columnwidth}}

\begin{figure}[t] 
\subfigure{\usebox{\tempbox}}
\subfigure{
\vbox to \ht\tempbox{ 
\vfil 
\epsfig{file=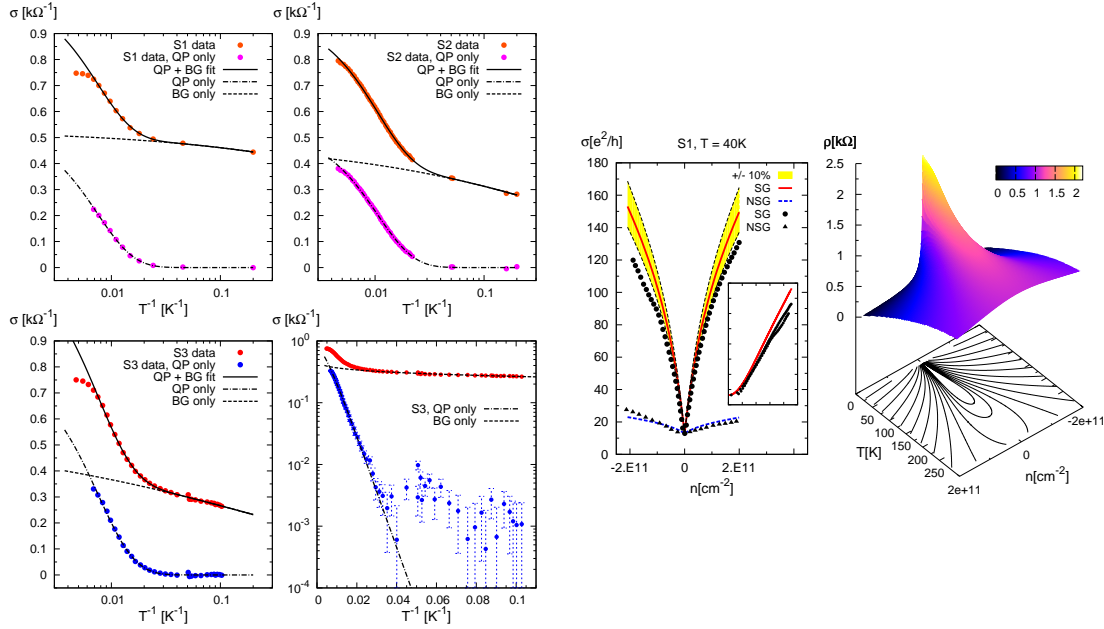, width = .48\columnwidth}
\vfil}}
\caption{Left panel: quasiparticle~(QP) and background~(BG) components of $\sigma(n=0,T)$ for the suspended 
graphene devices S1--S3, as determined in Ref.~\cite{DL_gap}. The empirical data is reproduced from Ref.~\cite{Bolotin_SG}. 
All devices show a ``knee" separating thermally activated and background regions. Right panel: $\sigma(n,T)$ determined from 
a fit to $\sigma(n=0,T)$ and the resistivity $\rho(n,T)$, reproduced from Ref.~\cite{DL_gap}.
\label{Fig_sigma}}
\end{figure}


The inclusion of the background component $\sigma_{bg}^{}$ is motivated by the fact that the minimal conductivity in graphene samples
is non-universal. In suspended graphene, it is much smaller than in graphene samples on a substrate and furthermore strongly 
sample-dependent. There is also a clear tendency of the minimal conductivity to decrease with increasing sample purity~\cite{Andrei_SG}.
It is likely that the minimal conductivity in graphene is formed of several components, including the effects of impurities~\cite{Herbut_disorder},
inhomogeneity~\cite{puddles} and effects due to invasive metallic contacts~\cite{Geim_contacts}. In our analysis, we have used
the phenomenological form
\beq
\label{sigma_bg}
\sigma_{bg}^{} \equiv \sigma_0^{} \, \exp[- (T_0^{}/T)^\alpha_{}], 
\eeq
which allows for the slight empirical $T$-dependence of $\sigma_{bg}^{}$. The empirical data of Ref.~\cite{Bolotin_SG} is shown
in Fig.~\ref{Fig_sigma}, together with fits in terms of $\sigma_q^{}$ and $\sigma_{bg}^{}$. It is noteworthy that the data display
a distinct ``knee'' at $\sim 30$~K, which in terms of the present description is interpreted as the temperature below which thermal
activation is negligible. Thus, in order to determine $\sigma_{bg}^{}$ in an unbiased fashion, we first fix $\sigma_0^{}$ and $T_0^{}$ using 
data in the extreme low-$T$ region. One may then subtract $\sigma_{bg}^{}$ at all $T$, and determine $\beta\eta$ and $\Delta$ by 
fitting $\sigma_q^{}$ to the resulting dataset. A simultaneous fit of $\sigma_q^{}$ and $\sigma_{bg}^{}$ confirms the validity of this
procedure. While $\eta(T)$ is {\it a priori} unknown, a scenario of constant $\beta\eta$ is strongly favored by the available data in 
the range $35$~K~$ \leq T \leq 150$~K.

Our findings in Ref.~\cite{DL_gap} suggest that the suspended graphene devices of Ref.~\cite{Bolotin_SG} exhibit a thermally 
activated conductivity $\sigma_q^{}$, which is well described by Eq.~(\ref{qp}) from $T \sim 150$~K down to $T \sim 35$~K, where 
the signal is lost due to limited measurement accuracy. The determined bandgaps are in the range $\Delta \sim 25 - 40$~meV, whereas
all samples were found to favor $\beta \eta \simeq 0.1$ indicating a scattering rate which increases linearly with $T$. A natural 
scattering mechanism with such properties is provided by the long-range Coulomb interaction~\cite{Fritz_cond} up to logarithmic corrections.
Furthermore, this value of $\beta\eta$ is consistent with the high carrier mobilities and long mean free paths reported in 
Ref.~\cite{Bolotin_SG}, as well as with theoretical estimates of the mean free path due to long-range Coulomb scattering~\cite{Fritz_cond}.
Specifically, for $T=35 - 150$~K we find $\eta=3.5 - 15$~K, with corresponding mean free paths of $\hbar v_F^{}/\eta \sim 0.5 - 2.0~\mu$m. 
Fits with no gap~($\Delta = 0$), constant~$\eta$, or zero background were found to be incompatible with data. 

As shown in Fig.~\ref{Fig_sigma}, these conclusions are consistent with the observed $\sigma(n)$, which depends sensitively on
the value of $\beta\eta$ determined from data at $n = 0$. Furthermore, the interpretation of the observed $\sigma(n,T)$ as due to
thermal activation accounts, in a natural way, for the observed transitional density $n^*_{}$ above which $\sigma(T)$ reverts from 
insulating to metallic. The determined scattering rate $\eta(T)$ is also suggestive of long-range Coulomb scattering, which is consistent
with the ultrapure character of the suspended graphene samples of Ref.~\cite{Bolotin_SG}.


\section{Conclusions}

We have reviewed the Lattice Gauge Theory approach to the low-energy EFT of graphene, with the aim of introducing
this technique to a wider audience and motivate the application of this approach to systems beyond monolayer graphene. Our 
calculations within this the graphene EFT indicate that it displays a chiral phase transition at a critical coupling of $\beta_c^{} = 0.0774(2)$,
with critical exponents that appear consistent with mean-field theory. Spontaneous chiral symmetry breaking in the graphene EFT would
lead to the appearance of a gap in the quasiparticle spectrum, directly linked to the formation of quasiparticle-hole pairs (excitons). It is
conceivable that such a transition occurs in suspended graphene, where the strength of the Coulomb interaction attains its maximum
value.

In an effort to clarify whether currently available experimental data on suspended graphene provide any evidence for the excitonic scenario,
we have reviewed the status of such measurements, which show a definite (though relatively mild) insulating trend at low~$T$ in the vicinity
of the neutral point. We have presented an interpretation of the observed anomalous temperature dependence in terms of the 
excitonic gap scenario, and tentatively found that the data may be consistently explained in terms of gapped Dirac 
quasiparticles ($\Delta \sim 30$~meV) with the long-range Coulomb interaction as a natural candidate for the dominant scattering 
mechanism. Further experimental studies of the conductivity at low~$T$ in suspended graphene are clearly called for, preferably minimizing 
the effects of invasive metallic contacts.

Further investigations using the Lattice Gauge Theory approach are in progress, including the renormalization of $v_F^{}$ due to the
Coulomb interaction, the magnetic catalysis of a semimetal-insulator transition at large external magnetic fields~\cite{Kh_Leal,Gorbar}, 
and the critical temperature for exciton condensation in graphene bilayers~\cite{graphene_bilayer}. The computation of transport 
properties involves the extraction of spectral functions in Minkowski spacetime from Euclidean time lattice data. Such calculations are also
feasible nowadays, as Bayesian analysis coupled with the Maximum Entropy Method has been successfully applied to 
QCD~\cite{graphene_conductivity}. Areas of interest include the electrical conductivity and viscosity~\cite{graphene_viscosity} of 
the quasiparticles
in graphene. Due to the flexibility of the Lattice Monte Carlo and Quantum Monte Carlo methods, further applications are likely to arise in
the closely related fields of graphene nanowires, high-$T_c^{}$ superconductors and hexagonal optical lattices, to name a few. In summary,
the application of Lattice Gauge Theory to condensed matter problems appears poised to develop into a highly fruitful field of study.



\end{document}